\documentclass[twocolumn, tighten]{aastex61}

\newcommand{\um}{$\mu$m}
\DeclareRobustCommand{\rchi}{{\mathpalette\irchi\relax}}
\newcommand{\irchi}[2]{\raisebox{\depth}{$#1\chi$}} 

\shorttitle{2MASS~J11151597+1937266: A Young, Dusty, Isolated Planetary-Mass Object}
\shortauthors{Theissen et al.}

\begin{document}

\title{2MASS~J11151597+1937266: A Young, Dusty, Isolated, Planetary-Mass Object with a Potential Wide Stellar Companion}

\author{Christopher A. Theissen}
\affiliation{Department of Astronomy, Boston University, 725 Commonwealth Avenue, Boston, MA 02215, USA}
\email{ctheisse@bu.edu}
\affiliation{Center for Astrophysics and Space Sciences, University of California, San Diego, 9500 Gilman Dr., Mail Code 0424, La Jolla, CA 92093, USA}

\author{Adam J. Burgasser}
\affiliation{Center for Astrophysics and Space Sciences, University of California, San Diego, 9500 Gilman Dr., Mail Code 0424, La Jolla, CA 92093, USA}

\author{Daniella C. Bardalez Gagliuffi}
\affiliation{Department of Astrophysics, American Museum of Natural History, Central Park West at 79th Street, New York, NY 10034, USA}

\author{Kevin K. Hardegree-Ullman}
\affiliation{The University of Toledo, 2801 West Bancroft Street, Mailstop 111, Toledo, OH 43606, USA}

\author{Jonathan Gagn\'e}
\affiliation{Carnegie Institution of Washington DTM, 5241 Broad Branch Road NW, Washington, DC 20015, USA}
\affiliation{NASA Sagan Fellow}

\author{Sarah J. Schmidt}
\affiliation{Leibniz-Institute for Astrophysics Potsdam (AIP), An der Sternwarte 16, D-14482, Potsdam, Germany}

\author{Andrew A. West}

\begin{abstract}

	We present 2MASS~J11151597+1937266, a recently identified low-surface gravity L dwarf, classified as an L2$\gamma$ based on Sloan Digital Sky Survey optical spectroscopy. We confirm this spectral type with near-infrared spectroscopy, which provides further evidence that 2MASS~J11151597+1937266 is a low-surface gravity L dwarf. This object also shows significant excess mid-infrared flux, indicative of circumstellar material; and its strong H$\alpha$ emission (EW$_{\mathrm{H}\alpha}=560\pm82$ \AA) is an indicator of enhanced magnetic activity or weak accretion. Comparison of its spectral energy distribution to model photospheres yields an effective temperature of $1724^{+184}_{-38}$ K. We also provide a revised distance estimate of $37\pm6$ pc using a spectral type--luminosity relationship for low-surface gravity objects. The 3-dimensional galactic velocities and positions of 2MASS~J11151597+1937266 do not match any known young association or moving group. Assuming a probable age in the range of 5--45 Myr, the model-dependent estimated mass of this object is between 7--21 $M_\mathrm{Jup}$, making it a potentially isolated planetary-mass object. We also identify a candidate co-moving, young stellar companion, 2MASS~J11131089+2110086. 

\end{abstract}


\section{Introduction}\label{intro}

	Young associations, such as nearby young moving groups (NYMGs) and open clusters, provide important benchmarks for testing stellar and brown dwarf evolutionary models \citep{zuckerman:2004:685}. There are a growing number of low-mass stars and brown dwarfs that show signatures of low-surface gravity and youth (ages $<100$~Myr), but are not associated with any currently known groups of young objects \citep[e.g.,][]{gagne:2015:33}. These very low-mass isolated objects are challenging to characterize due to the difficulty in precisely constraining their ages, a necessary step to break the mass-age-temperature degeneracy for brown dwarfs. They may indicate new associations still awaiting discovery, or evidence of brown dwarf ejection from clusters \citep{boss:2001:l167,hoogerwerf:2001:49,reipurth:2001:432,bate:2002:l65}. 

	\citet{theissen:2017:92} recently identified 2MASS J11151597+1937266 (hereafter 2MASS~J1115+1937) in the Late-Type Extension to the Motion Verified Red Stars catalog (LaTE-MoVeRS) as a very-low-mass, ultracool object (spectral type L2; $T_\mathrm{eff} \approx 1700$ K), with signatures of either accretion or a flaring event based on strong H and He optical line emission. 2MASS~J1115+1937 also shows significant excess mid-infrared (MIR) flux, which may be indicative of primordial circumstellar material \citep[e.g.,][]{faherty:2013:2}. In this study, we present evidence that 2MASS~J1115+1937 is likely a young ($\lesssim45$ Myr), potentially planetary-mass object ($\lesssim13~M_\mathrm{Jup}$) whose kinematics are inconsistent with any known young association. We also discuss a candidate co-moving stellar companion that also shows signatures of youth, 2MASS~J11131089+2110086 (hereafter 2MASS~J1113+2110).

\section{Characterization of 2MASS~J1115+1937}
\label{characterizing}

\subsection{Spectral Typing}

\begin{figure*}
\centering
\includegraphics[width=\textwidth]{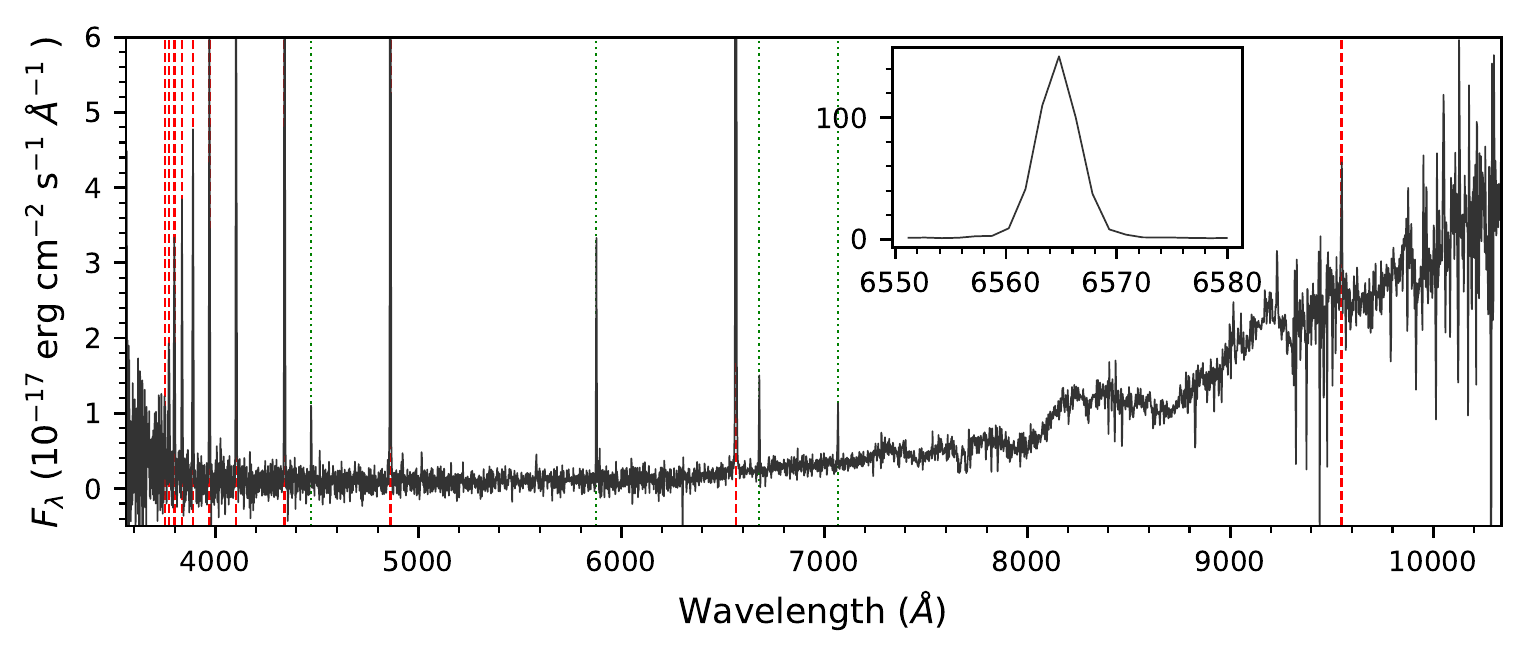}
\caption{SDSS spectrum of 2MASS~J1115+1937. Red dashed lines denote hydrogen transitions, and green dotted lines denote helium transitions. The y-scale has been truncated to show more spectral features, thus clipping many of the hydrogen emission lines. The inset plot shows the H$\alpha$ emission line.
\label{fig:sdss}}
\end{figure*}

	\citet{theissen:2017:92} used the optical spectrum from the Sloan Digital Sky Survey \citep[SDSS;][]{york:2000:1579} Baryon Oscillation Spectroscopic Survey \citep[BOSS;][]{dawson:2013:10} to spectral type 2MASS~J1115+1937 as L2 (see Figure~\ref{fig:sdss} below and Figure 14 from \citealt{theissen:2017:92}). However, the significant veiling in the continuum, possibly due to accretion \citep{white:2003:1109} or a flaring event \citep{kowalski:2013:15}, made optical spectral typing difficult. The best visual match to the optical spectrum was found to be the low-surface gravity L2 dwarf 2MASS~J23225299$-$6151275 \citep{cruz:2009:3345}, particularly at the redder end of the spectrum ($> 8000$~\AA). We measured a radial velocity (RV) for 2MASS~J1115+1937 of $-14\pm7$ km s$^{-1}$ by simultaneously fitting Gaussian functions to all of the hydrogen lines with a Markov Chain Monte Carlo method built using the \textit{emcee} code \citep{foreman-mackey:2013:306}.
	
	We obtained a low-resolution near-infrared (NIR) spectrum of 2MASS~J1115+1937 using the prism-dispersed mode of the SpeX spectrograph \citep{rayner:2003:362} on the NASA Infrared Research Telescope Facility (IRTF) on 2017 May 6 (UT). Using the 0$\farcs$5 slit, we obtained spectra with a resolution of $\approx$120 over a wavelength range of 0.8--2.5~\um. The SpeX data were reduced using the SpeXtool package \citep{vacca:2003:389, cushing:2004:362} following standard procedures. The resulting spectrum was analyzed using the SpeX Prism Library Analysis Toolkit\footnote{\url{http://www.browndwarfs.org/splat}.} (SPLAT; Burgasser et al., in preparation) and spectral templates from the SpeX Prism Library \citep[SPL;][]{burgasser:2014:}.
	
\begin{figure*}
\centering
\includegraphics[width=.496\textwidth]{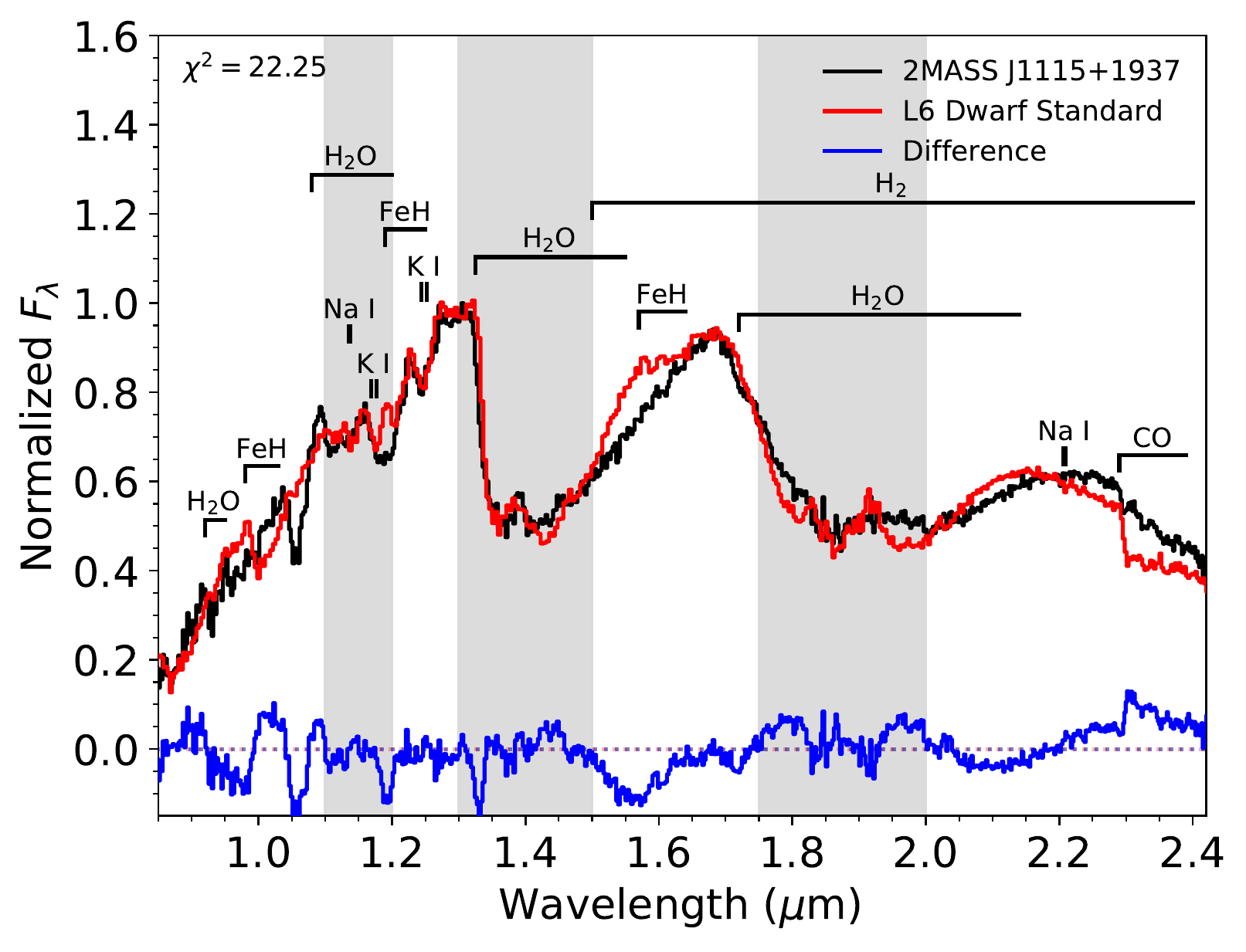}
\includegraphics[width=.496\textwidth]{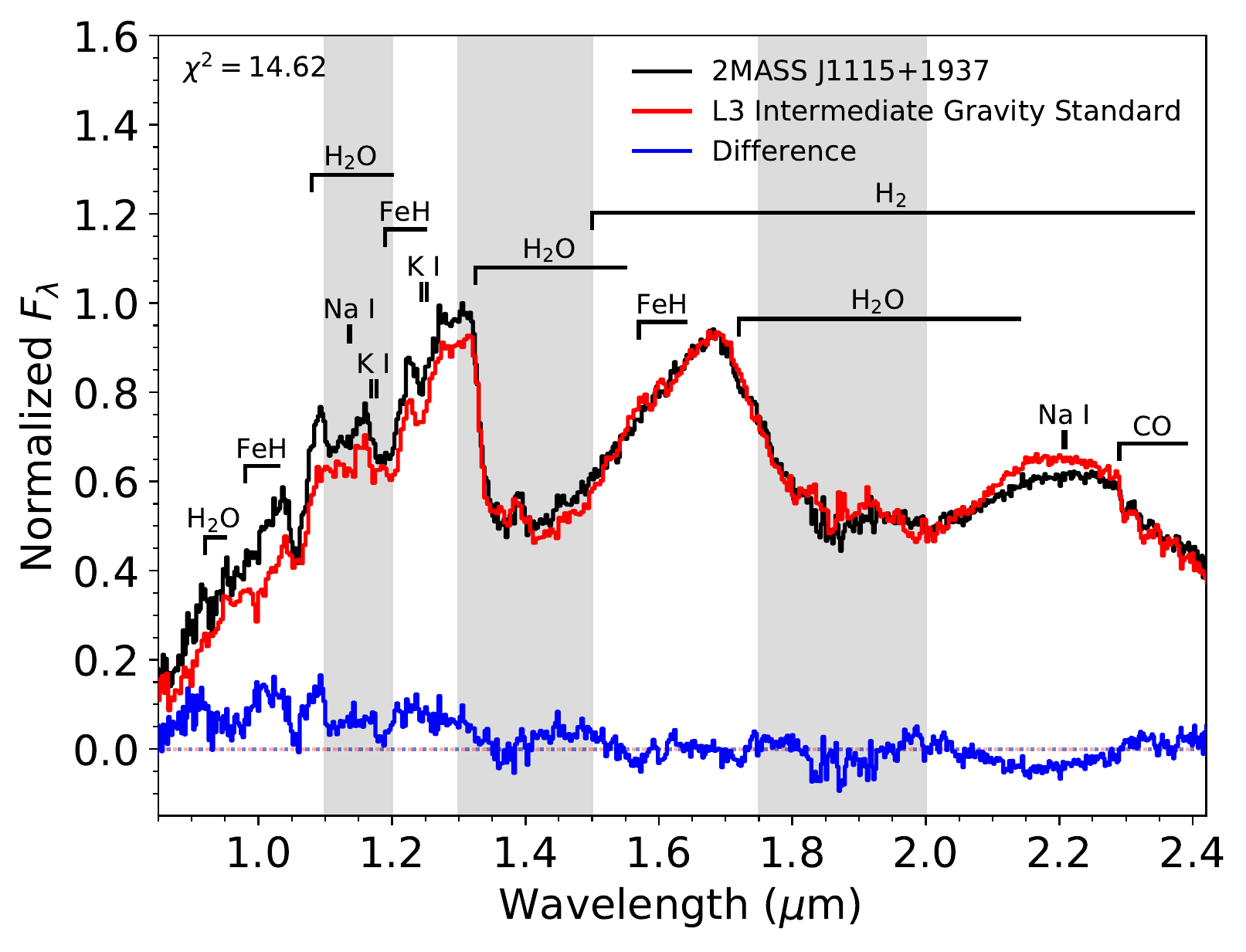}
\includegraphics[width=.496\textwidth]{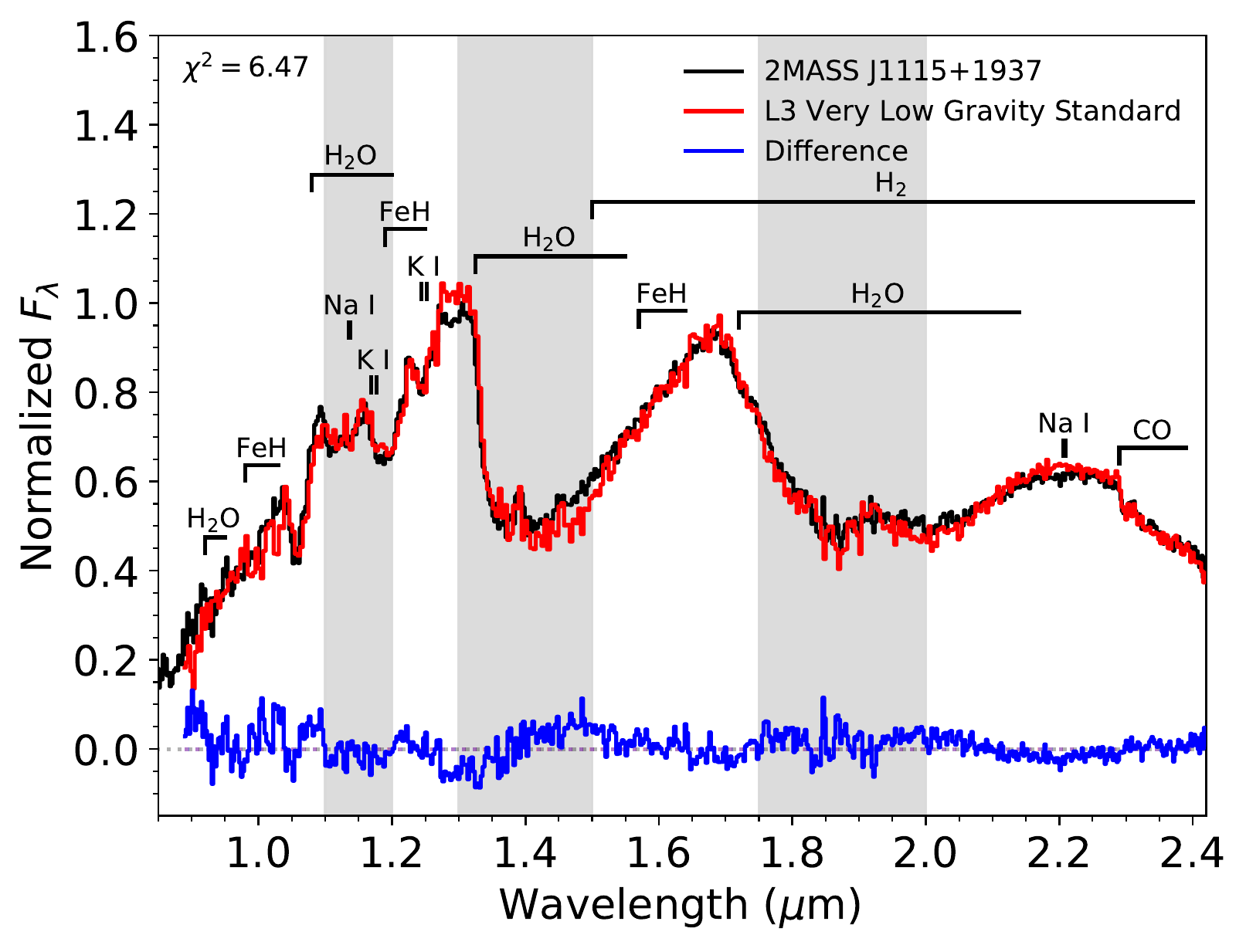}
\includegraphics[width=.496\textwidth]{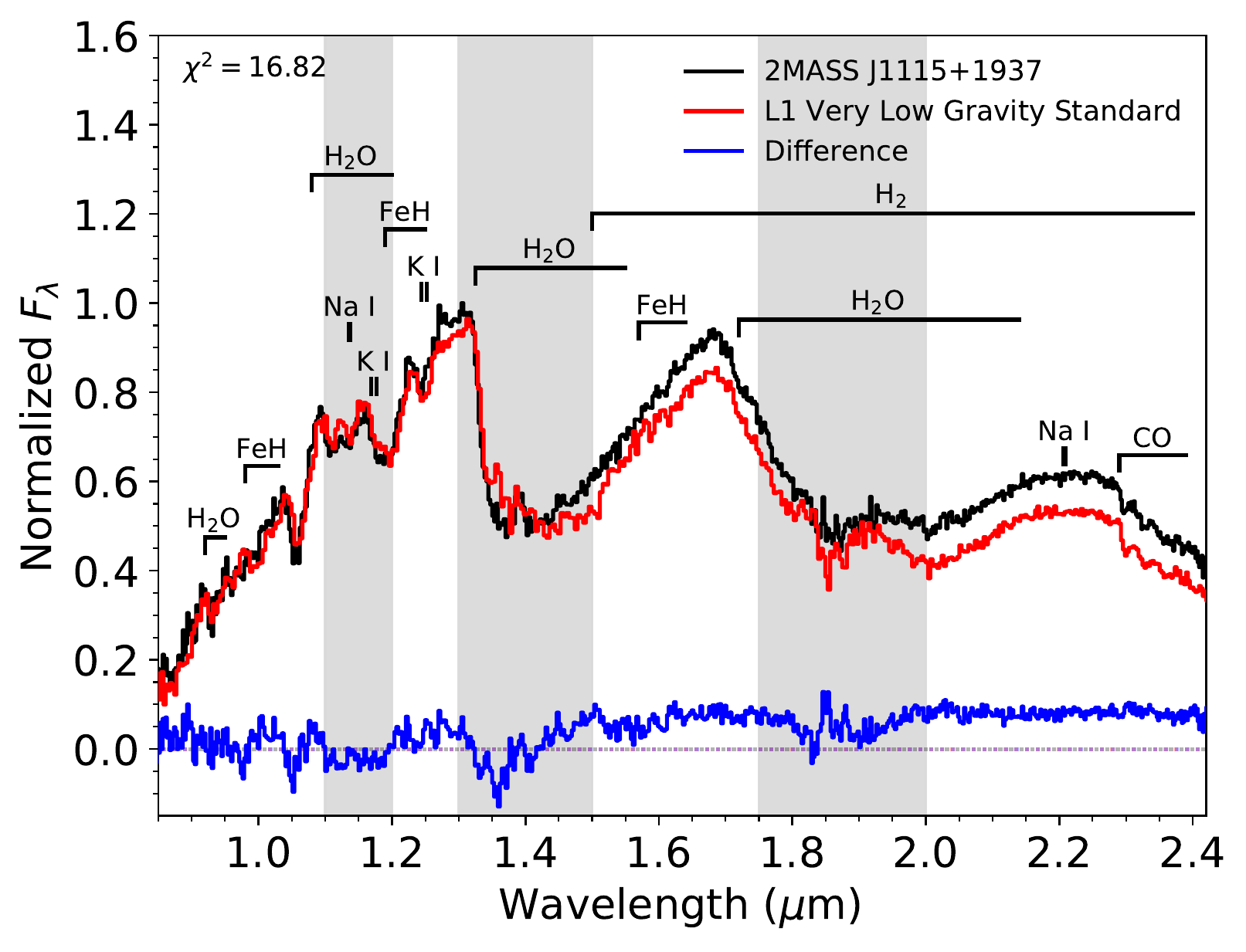}
\caption{Comparisons between 2MASS~J1115+1937 (black lines) and the best-fit spectral standards at various gravities (red lines). The difference spectra are shown in blue. The gray dotted horizontal line indicates a difference of zero. The gray shaded areas indicate telluric regions, and absorption features are labeled.
\textit{Top Left}: Comparison to the L6 (field gravity) standard 2MASS~J1010148$-$0406499 \citep[data from][]{reid:2006:1114}.
\textit{Top Right}: Comparison to the L3$\beta$ standard 2MASS~J1726000$+$1538190 \citep[data from][]{allers:2013:79}
\textit{Bottom Left}: Comparison to the L3$\gamma$ 2MASS~J22081363$+$2921215 \citep[data from][]{allers:2013:79}, a field object with an estimated age of 20--26~Myr \citep{gagne:2015:33}.
\textit{Bottom Right}: Comparison to the L1$\gamma$ 2MASS~J05184616$-$2756457 \citep[data from][]{allers:2013:79}. This object represents the closest NIR spectral type using the spectral-typing schemas of \citet{allers:2013:79} and Cruz et al. (2017, \textit{AJ}, submitted). This final comparison was done using only the 0.9--1.4~\um\ region \citep{kirkpatrick:2010:100}. The overall best statistical fit to 2MASS~J1115+1937 using the entire spectrum is 2MASS~J22081363$+$2921215.
\label{fig:nir}}
\end{figure*}

	Figure~\ref{fig:nir} shows the best-fit comparisons to spectral standards among field dwarfs, intermediate surface gravity dwarfs ($\beta$), very-low-surface gravity dwarfs ($\gamma$), and extremely-low-surface gravity dwarf ($\delta$), proposed in \citet{kirkpatrick:2005:195} with templates defined in \citet{cruz:2009:3345}. These gravity classifications, $\beta$, $\gamma$, and $\delta$, roughly coincide with $\log g$ ranges of 4--4.5, 3.5--4 and 3--3.5 (cgs), respectively, although \citet{allers:2013:79}, \citet{gagne:2015:33}, and \citet{martin:2017:73} have shown that gravity classifications have considerable scatter with respect to cluster ages, which in turn impacts the parameters inferred from model atmosphere fits. These gravity classifications are useful only as relative measures of age/gravity within a spectral type.

	The best match to a sample of field objects is the L6 dwarf standard 2MASS~J1010148$-$0406499, which is a considerably later type than the optical classification. The best statistical fit using a broader template catalog (using the $\rchi^2$ method described in \citealt{bardalez-gagliuffi:2014:143}) is to the L3$\gamma$ 2MASS~J22081363$+$2921215 \citep{allers:2013:79}, which is a member of the 20--26~Myr old $\beta$ Pictoris group \citep{gagne:2015:33}. 

	Using the index-based classification scheme from \citet{allers:2013:79}, we obtained a NIR spectral type of L1 and a gravity classification of very low gravity (VL-G, equivalent to $\gamma$) based on the FeH and VO band strengths, alkali line depths, and triangular shape of the $H$-band continuum. A comparison to the L1$\gamma$ standard 2MASS~J05184616$-$2756457 \citep{allers:2013:79} is shown in Figure~\ref{fig:nir}. A spectral type of L1$\gamma$ was also found using the method of normalizing and comparing the NIR spectrum band-by-band ($zJ/H/K$; Cruz et al. 2017, \textit{AJ}, submitted). Combining these analyses, we adopt a mean NIR spectral-type of L2$\gamma$ ($\pm1$), which is consistent with the optical classification.

\subsection{Constructing the Spectral Energy Distribution}

\begin{figure*}
\centering
\includegraphics{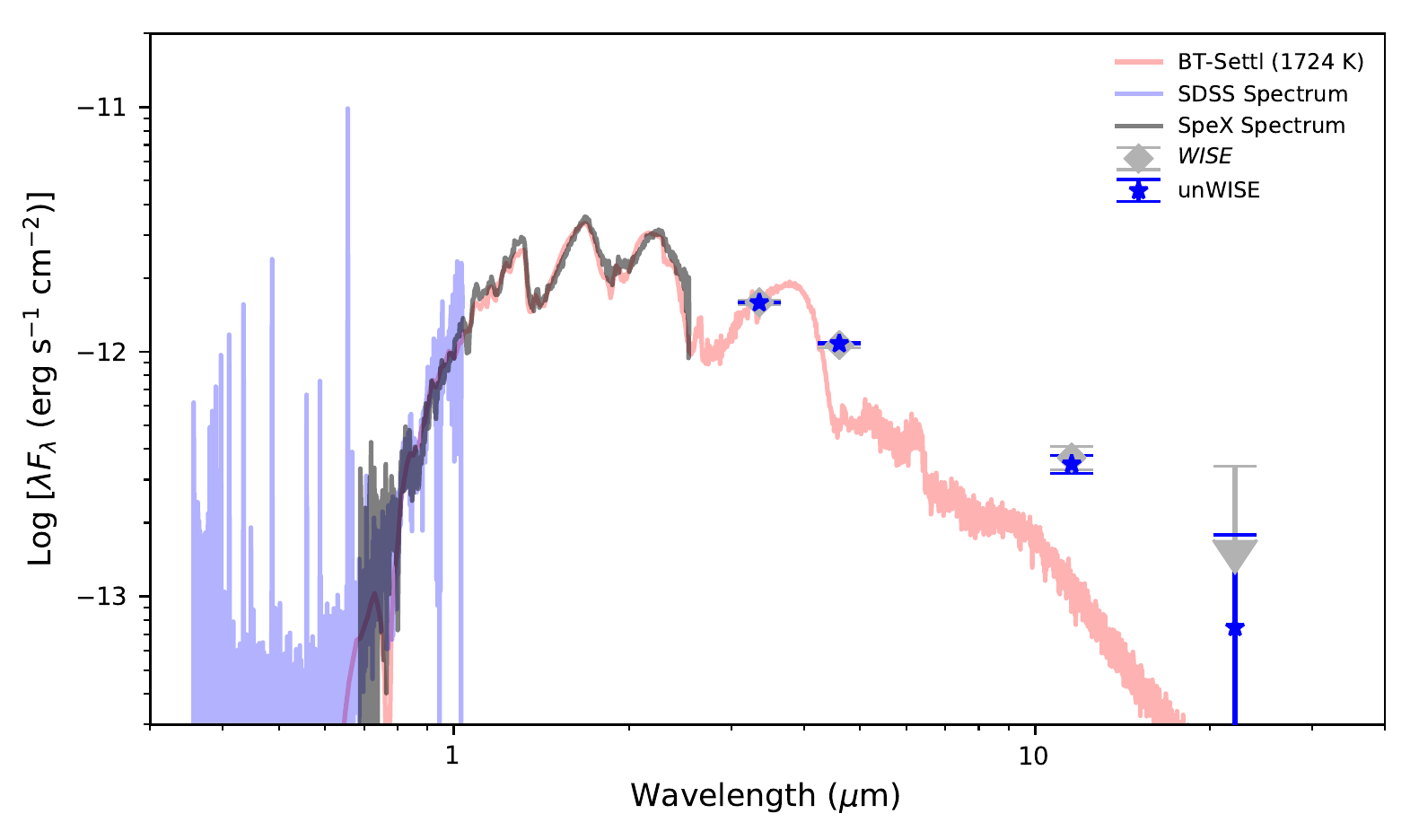}
\includegraphics{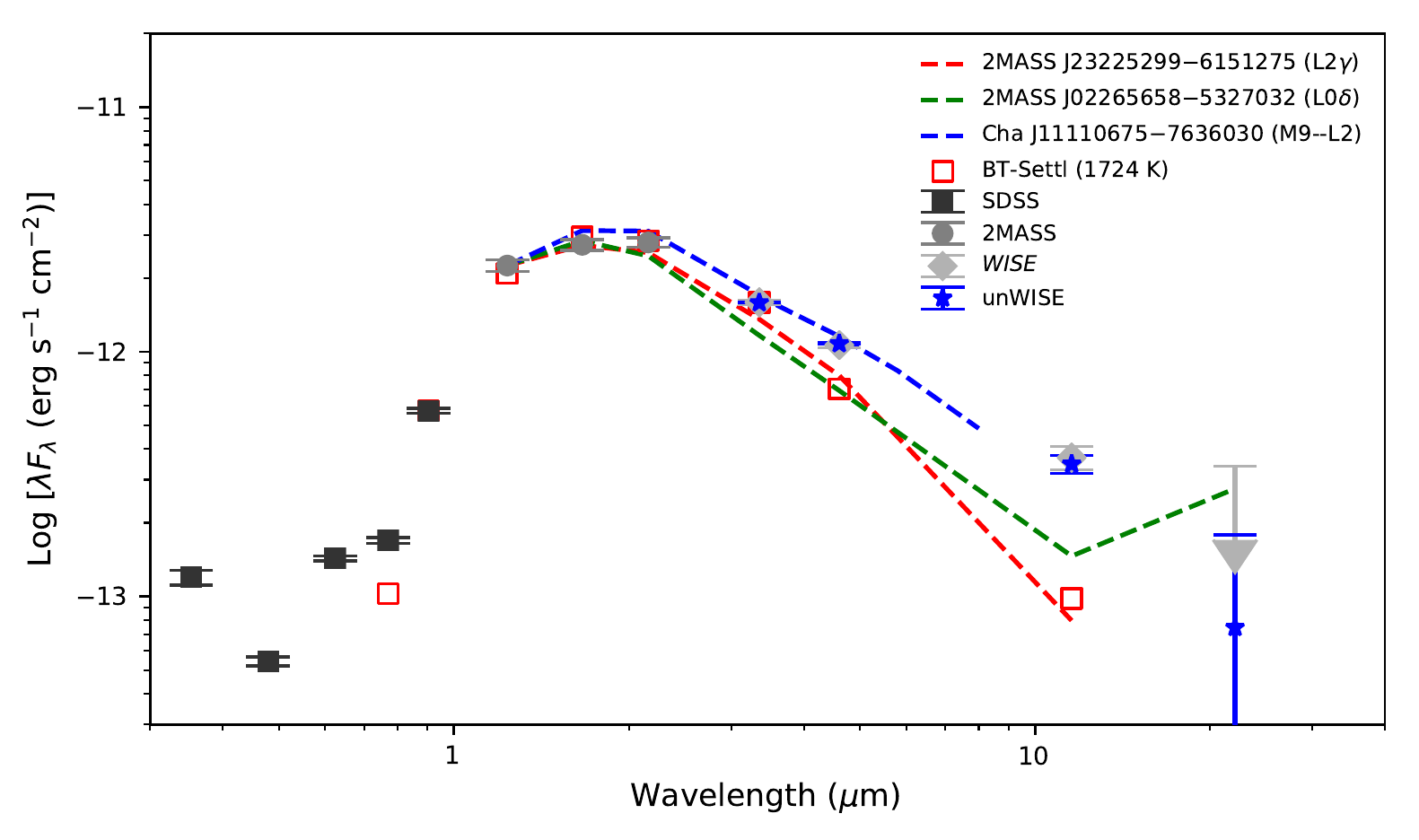}
\caption{SEDs for 2MASS~J1115+1937. 
\textit{Top:} Spectra from SDSS (blue line) and SpeX (gray line) along with the best-fit BT-Settl model (red line). Measurements from AllWISE (light gray diamonds; arrows indicate upper limits) and unWISE (blue stars) are also shown.
\textit{Bottom:} Measurements from SDSS (black squares) and 2MASS (gray circles). Expected bandpass integrated fluxes from the best-fit BT-Settl model are shown as red squares. The SEDs for 2MASS~J02265658$-$5327032 (green dashed line) and 2MASS~J22081363+2921215 (red dashed line) are shown to highlight the excess MIR flux of 2MASS~J1115+1937. Also shown is the young (1--3 Myr) brown dwarf Cha J11110675$-$7636030 (blue dashed line) which has been shown to exhibit a MIR excess \citep{esplin:2017:46}.
\label{fig:sed}}
\end{figure*}

	Broad-band spectral energy distributions (SEDs) can constrain the effective temperatures of very low-mass dwarfs and reveal evidence of MIR excesses. Photometry from SDSS, 2MASS, and \textit{WISE} for 2MASS~J1115+1937 is shown in Figure~\ref{fig:sed}. We re-fit the $zJHK_sW1$ photometry to the BT-Settl model photospheres \citep{allard:2012:2765, allard:2013:128} using a Markov Chain Monte Carlo (MCMC) routine described in \citet{theissen:2017:92} and \citet{theissen:2017:165}. The reconstructed SED is shown in Figure~\ref{fig:sed}, and the photometry and model values are listed in Table~\ref{tbl:candidate}. We include the forced \textit{WISE} photometry measurements (photometry forced at the SDSS source position within the \textit{WISE} images) from the unWISE coadds \citep{lang:2014:108}, which typically have better noise estimates than conventional \textit{WISE} photometry \citep{lang:2014:108, lang:2016:36}.

\clearpage

	The SED of 2MASS~J1115+1937 shows elevated fluxes in the near-UV/optical (SDSS-bands) and the MIR. The higher levels of MIR flux may be due to dust in the system, similar to what \citet{boucher:2016:50} discovered for 2MASS~J02265658$-$5327032, an L0$\delta$ dwarf and possible member of the $\sim$40~Myr Tucana-Horologium association \citep{kraus:2014:146}. As previously noted, the elevated near-UV and optical flux levels, and optical line emission, may be signs of accretion or flaring. These elevated flux levels are persistent, as they are present in SDSS photometry and spectroscopy taken at different epochs.
	
	We made comparisons to known, young brown dwarfs (Figure~\ref{fig:sed}), all scaled to the 2MASS~J1115$+$1937 $J$-band flux. 2MASS~J02265658$-$5327032 \citep[L0$\delta$;][]{gagne:2015:33} and 2MASS~J23225299$-$6151275 \citep[L2$\gamma$;][]{cruz:2009:3345} are both low-surface gravity dwarfs with age estimates $<50$~Myr \citep{gagne:2015:33}. 2MASS~J02265658$-$5327032 is one of the oldest and lowest mass objects known still harboring a primordial disk based on excess MIR flux beyond 10~\um\ (\textit{WISE} $W3$ and $W4$). Figure~\ref{fig:sed} (bottom) also shows the SED for Cha J11110675$-$7636030, a potential planetary-mass object (3--10~$M_\mathrm{Jup}$) in the 1--3~Myr Chamaeleon I star-forming region \citep{esplin:2017:46}. Cha J11110675$-$7636030 has the most similar SED to 2MASS~J1115+1937, possibly indicating that 2MASS~J1115+1937 is also extremely young. 
	 
	 \citet{esplin:2017:46} discuss the potential for Cha J11110675$-$7636030 to be the least massive known brown dwarf (3--6 $M_\mathrm{Jup}$) hosting a primordial circumstellar disk, although further additional data are needed to confirm the MIR excess arises from a circumstellar disk. Although there is not enough evidence to support 2MASS~J1115+1937 having an age between 1--3~Myr, we can assume a lower age limit of 5 Myr, coinciding with the end of the accretion phase for brown dwarfs through observations of hydrogen and helium emission \citep{mohanty:2005:498}. We can also assume an upper age limit of 45~Myr, which corresponds to the oldest known objects harboring primordial circumstellar material and potentially accreting \citep{boucher:2016:50, murphy:2017:}.

\startlongtable
\begin{deluxetable*}{lll}
\tabletypesize{\footnotesize}
\tablecolumns{3}
\tablecaption{Object Properties\label{tbl:candidate}}
\tablehead{
\colhead{Parameter} & \colhead{2MASS~J1115+1937} & \colhead{2MASS~J1113+2110}
}
\startdata
SDSS DR8$+$ objID							& 1237667915950588237					& 1237667734502047961\\
R.A. (deg.)								& $168.816447$						& $168.295285$\\
Decl.	 (deg.)								& $19.624012$							& $21.169016$\\
Spectral type ($\pm1$)						& L2$\gamma$ (optical)					& M6 (optical)\\
										& L2$\gamma$ (NIR)					& M7 (NIR)\\
$T_\mathrm{eff}$ (K)							& $1724^{+184}_{-38}$ 					& $2767^{+42}_{-53}$ \\
$d_\mathrm{phot}$ (pc)						& $37\pm6$ ($48\pm6$)\tablenotemark{a}		& $54\pm9$ \\
RV (km s$^{-1}$)							& $-14\pm7$ 			 				& $-10.1\pm0.3$ \\
$v \sin i$ (km s$^{-1}$)						& ... 									& $15.2\pm1.3$ \\
$\mu_\alpha$ $\cos \delta$ (mas yr$^{-1}$)\tablenotemark{b}			& $-57\pm13$ 							& $-67\pm8$\\
$\mu_\delta$ (mas yr$^{-1}$)\tablenotemark{b}						& $-25\pm8$ 							& $-14\pm11$ \\ 
$U$ (km s$^{-1}$)\tablenotemark{c,d}			& $-3\pm3$							& $-10\pm3$\\
$V$ (km s$^{-1}$)\tablenotemark{d}				& $-3\pm3$							& $-6\pm2$\\
$W$ (km s$^{-1}$)\tablenotemark{d}				& $-17\pm7$							& $-16\pm1$ \\
$X$ (pc)									& $-10\pm2$ 							& $-16\pm3$ \\
$Y$ (pc)									& $-11\pm2$ 							& $-14\pm2$ \\
$Z$ (pc)									& $34\pm6$ 							& $50\pm8$ \\
Radius ($R_\odot$)\tablenotemark{e}			& $0.13\pm0.02$ 						& $0.14\pm0.02$\\
$J$ (2MASS)								& $15.56\pm0.06$						& $13.87\pm0.02$\\
$H$ (2MASS)								& $14.57\pm0.06$						& $13.24\pm0.03$\\
$K_s$ (2MASS)							& $13.80\pm0.05$						& $12.89\pm0.03$\\
$W1$ (AllWISE)							& $13.09\pm0.02$						& $12.65\pm0.02$\\
$W2$ (AllWISE)							& $12.55\pm0.03$						& $12.45\pm0.03$\\
$W3$ (AllWISE)							& $10.77\pm0.12$						& $12.23\pm0.41$\\
$W4$ (AllWISE)							& $> 8.68$							& $> 8.42$\\
$W1$ (unWISE)							& $13.095\pm0.003$						& $12.654\pm0.002$\\
$W2$ (unWISE)							& $12.533\pm0.007$						& $12.437\pm0.006$\\
$W3$ (unWISE)							& $10.840\pm0.093$						& $12.286\pm0.352$\\
$W4$ (unWISE)							& $10.330\pm1.509$						& $10.490\pm1.731$\\
EW$_{\mathrm{H}\alpha}$ (\AA)\tablenotemark{f}	& $560\pm82$							& $30\pm2$ \\
$\log (L_{\mathrm{H}\alpha} / L_\mathrm{bol})$		& $-2.9\pm0.1$							& $-3.27\pm0.03$ \\
$M_J$ 									& $12.75\pm0.36$\tablenotemark{g}			& $10.20\pm0.36$\\ 
$M_H$ 									& $11.73\pm0.36$\tablenotemark{g}			& $9.58\pm0.36$\\ 
$M_{K_s}$ 								& $10.96\pm0.36$\tablenotemark{g}			& $9.22\pm0.36$\\ 
Age (Myr)									& 5--45 								& $\sim$100? \\
Mass ($M_\odot$)							& 0.007--0.021 							& 0.043--0.072\tablenotemark{h} \\
$\log (L_\ast / L_\odot)$						& $-3.82\pm0.12$ 						& $-2.97\pm0.08$\\
\enddata
\tablenotetext{a}{Original distance value from \citet{theissen:2017:92}.}
\tablenotetext{b}{These values come from the LaTE-MoVeRS catalog.}
\tablenotetext{c}{Positive values indicate motion towards the Galactic center, ensuring that the $UVW$ frame of reference is a right-handed coordinate system.}
\tablenotetext{d}{These values have \textit{not} been corrected for Solar motion.}
\tablenotetext{e}{Based on the photometric distance value and best-fit BT-Settl model.}
\tablenotetext{f}{Positive values indicate H$\alpha$ emission.}
\tablenotetext{g}{Based on the ``young" calibration from \citet{faherty:2016:10} Table 19 and optical spectral types.}
\tablenotetext{h}{Assuming an age range of 45--100 Myr.}
\end{deluxetable*}

\subsection{Accretion or Flare?: Persistent Hydrogen Emission}
\label{hydrogen}

\begin{figure*}
\centering
\includegraphics{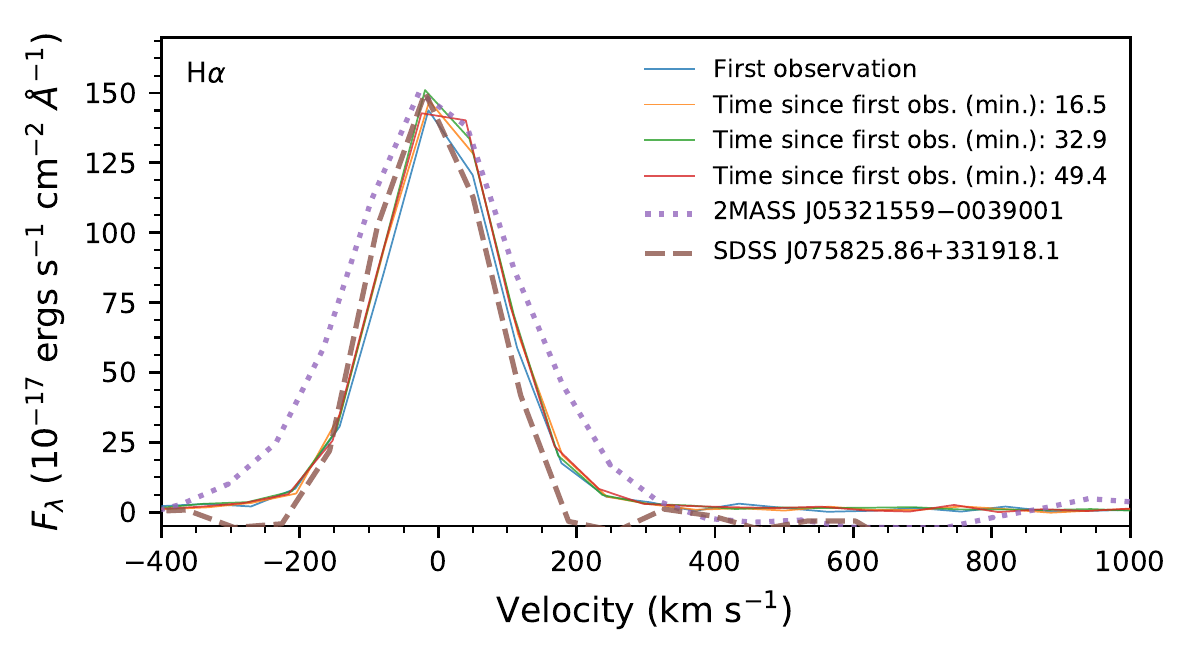}
\includegraphics{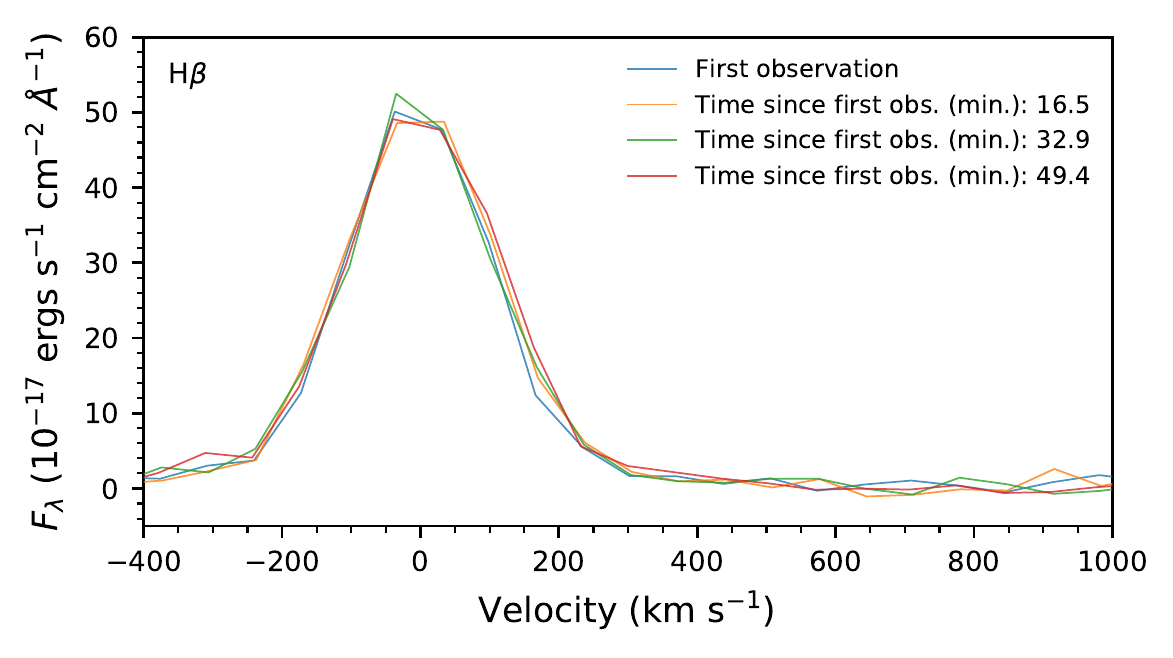}
\includegraphics{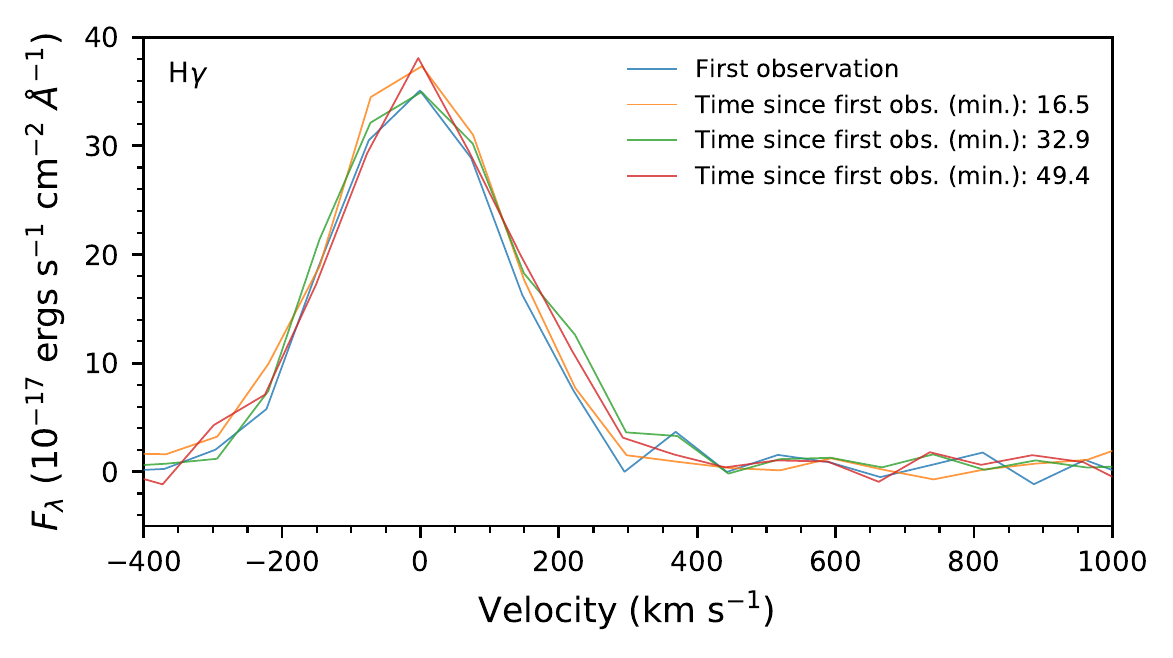}
\caption{Hydrogen emission line profiles (H$\alpha$, H$\beta$ and H$\gamma$) for each single SDSS observation of 2MASS~J1115+1937. There is no apparent increase or decrease in the line emission over the $\sim$50 minute period between the first and last observations. We show the scaled H$\alpha$ line profile for 2MASS~J05321559$-$0039001 (dotted line), a young ($\lesssim 1$ Myr) M4 dwarf in Orion with signs of accretion. We also show the scaled H$\alpha$ line profile for SDSS J075825.86$+$331918.1 (dashed line), an active (EW$_{\mathrm{H}\alpha}=40\pm9$~\AA) field M9 dwarf. The line profile for 2MASS~J1115+1937 is more similar to the field dwarf than the accreting object, likely indicating that the emission from 2MASS~J1115+1937 is not due to strong accretion.
\label{fig:hydrogen}}
\end{figure*}

	Extremely young ($\lesssim 1$~Myr), accreting objects have persistent (but variable) hydrogen emission \citep[variability on the order of days;][]{dupree:2012:73}, and broad emission lines \citep[e.g.,][]{white:2003:1109}. Flaring events produce narrow-line hydrogen emission with fluxes that evolve in a well-defined pattern \citep[e.g.,][]{hilton:2010:1402, kowalski:2013:15}. The SDSS spectrum is a co-added spectrum of four different spectra taken over a $\sim$50 minute period, as is shown in Figure~\ref{fig:hydrogen}. From \citet{kowalski:2013:15}, we can surmise that a flare observed close to peak emission will have an H$\alpha$ line flux that increases or decreases by about a factor of two over this time frame. This is not seen in Figure~\ref{fig:hydrogen}. Comparison to a known accreting $\sim$1~Myr M4 dwarf in Orion (2MASS~J05321559$-$0039001) shows a significantly narrower line profile than that expected from an accreting object. Comparison to the active field M9 dwarf SDSS J075825.86$+$331918.1 shows a similar line profile, with marginally broader wings for 2MASS~J1115+1937. \citet{jayawardhana:2003:282} note that very-low-mass accreting objects tend to have narrower line profiles as compared to low-mass stars. 2MASS~J1115+1937 does not appear to be strongly accreting, nor is the hydrogen emission likely related to a flaring event. 

	2MASS~J1115+1937 also shows He \textsc{i} (6678~\AA) emission (EW$_\mathrm{He~\textsc{i}}=25\pm10$~\AA), another signature of accretion \citep{mohanty:2005:498}. We deduce that the hydrogen emission is due to persistent, enhanced magnetic activity \citep{west:2015:3} and/or weak accretion \citep{mohanty:2005:498}, both signatures of youth. Using the temperature range from the MCMC (1686--1908 K) and assuming an age range of 5--45~Myr, we infer a mass range between 7--21~$M_\mathrm{Jup}$ estimated using the minimum and maximum ranges of the evolutionary models of \citet{burrows:2001:719}, \citet{baraffe:2003:701}, \citet{saumon:2008:1327}, and \citet{baraffe:2015:a42}.

\subsection{Revising the Distance}
\label{distance}

	\citet{theissen:2017:92} estimated a distance of $48\pm6$ pc for 2MASS~J1115+1937 using the photometric distance relationships from \citet{schmidt:2016:9999}. These relationships were calibrated using field objects from SDSS, and contain very few young/low-surface gravity objects. We revised the distance estimate using the absolute magnitude--spectral type relationships for ``young" objects from \citet{faherty:2016:10} Table 19, which use optical spectral types. For an adopted (optical) spectral type of L2, we obtained photometric distances $40\pm12$~pc, $38\pm11$~pc, $35\pm10$~pc using the $J$, $H$, and $K_s$ magnitudes, respectively, for an uncertainty weighted average distance of $37\pm6$~pc. This is 1.5-$\sigma$ closer than the previously estimated distance, and assumes 2MASS~J1115+1937 is a single object.

	Using a distance of $37\pm6$~pc gives a luminosity estimate of $\log(L_\ast / L_\odot) = -3.82\pm0.12$. This luminosity is more consistent with the expected value for a field L2 dwarf than a low gravity L2 dwarf \citep[see Figure 32 from][]{faherty:2016:10}. If we use the ``young" calibration from \citet{faherty:2016:10}, we obtain an expected value of $\log(L_\ast / L_\odot)-3.29\pm0.13$ for a low gravity L2 dwarf. The distance value of $48\pm6$~pc from \citet{theissen:2017:92}, gives a luminosity of $\log(L_\ast / L_\odot) = -3.59\pm0.12$, which is within the spread for ``young" L2 dwarfs \citep{faherty:2016:10}. This could indicate that the true distance to 2MASS~J1115+1937 is closer to the field calibration than the young calibration.

	Unfortunately 2MASS~J1115+1937 is not contained within \textit{Gaia} Data Release 1 \citep[DR1;][]{gaia-collaboration:2016:a2}, and will require future ground- or space-based observations to obtain a trigonometric parallax measurement to determine the true distance and luminosity of 2MASS~J1115+1937. In light of the numerous signatures of youth for 2MASS~J1115+1937, for the remainder of this study we assume a distance of $37\pm6$~pc.

\section{2MASS~J11131089+2110086: A Potentially Co-moving Star?}
\label{neighbor}
	
	We searched the LaTE-MoVeRS, MoVeRS \citep{theissen:2016:41}, and \textit{Gaia} DR1 catalogs for nearby, low-mass companions (that could be part of a new association) within 2$^\circ$ of 2MASS~J1115+1937 that exhibited similar proper motions and distances. The proper motion information listed in Table~\ref{tbl:candidate} for 2MASS~J1115+1937 comes from the LaTE-MoVeRS catalog. We found a potential co-moving star, 2MASS~J1113+2110, at an angular separation of 1.62$^\circ$ (corresponding to a physical separation of 1.05~pc at a distance of 37~pc). We obtained a moderate-resolution optical spectrum of this source using the DeVeny spectrograph \citep[$\lambda / \Delta \lambda \approx 2800$;][]{bida:2014:91472n} on the 4.3-m Discovery Channel Telescope on 2017 April 25 (UT), a low-resolution NIR spectrum using SpeX on 2017 May 6 (UT), and a high-resolution NIR spectrum ($\lambda/\Delta\lambda\approx20000$) on the Keck II 10-m telescope using the Near InfraRed Spectrometer \citep[NIRSPEC;][]{mclean:2000:1048} on 2017 May 5 (UT).
	
\begin{figure*}
\centering
\includegraphics[width=0.496\textwidth]{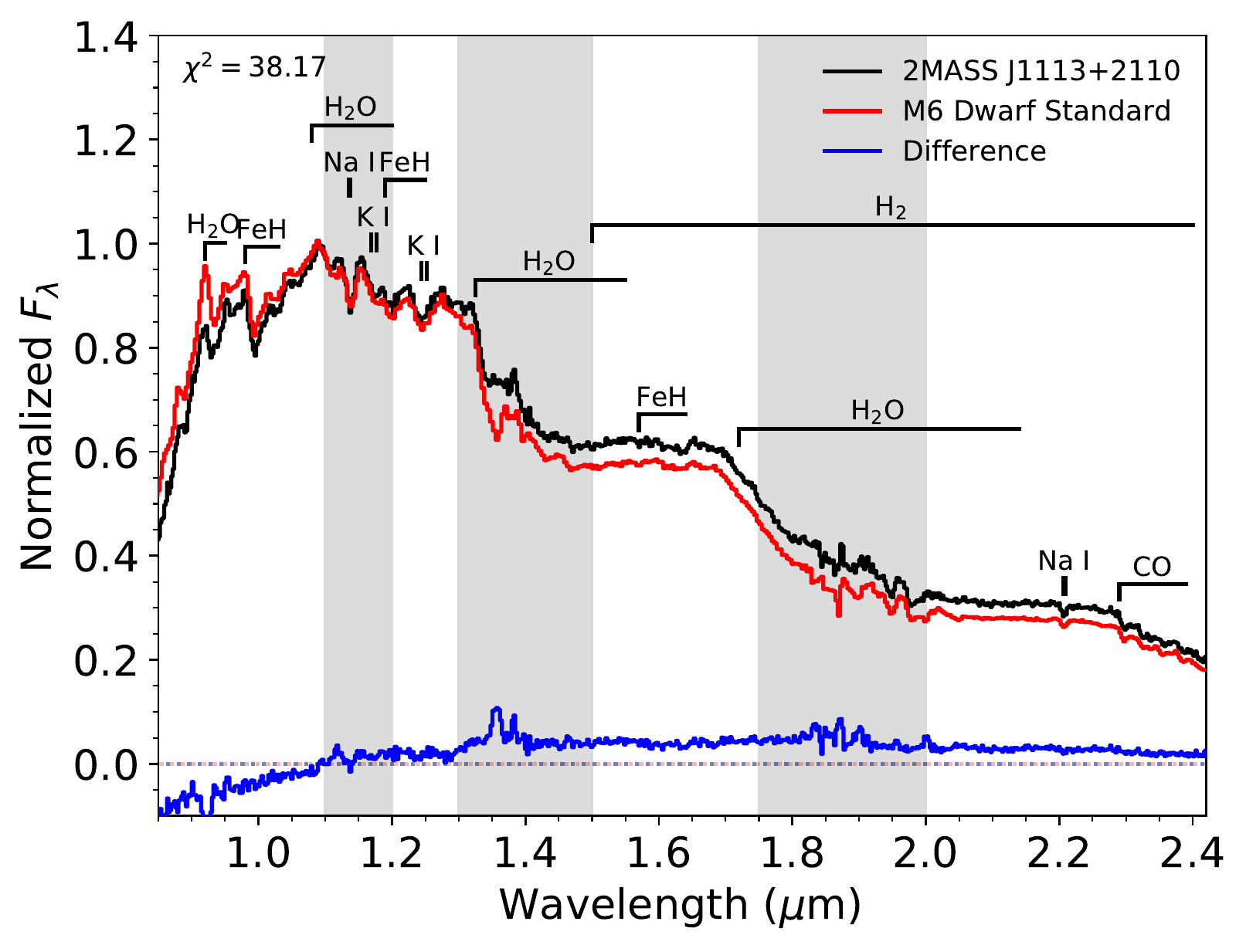}
\includegraphics[width=0.496\textwidth]{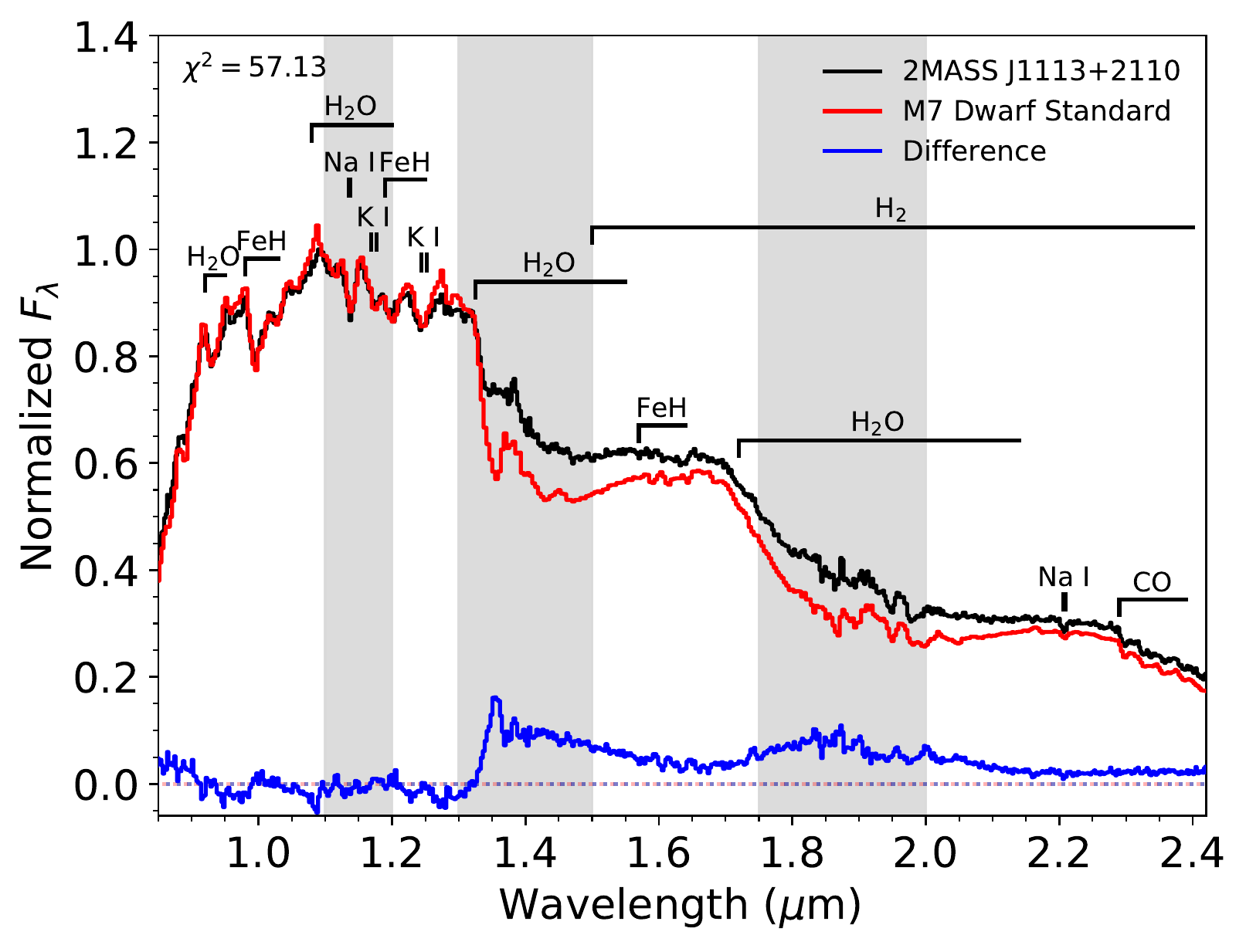}
\includegraphics[width=0.496\textwidth]{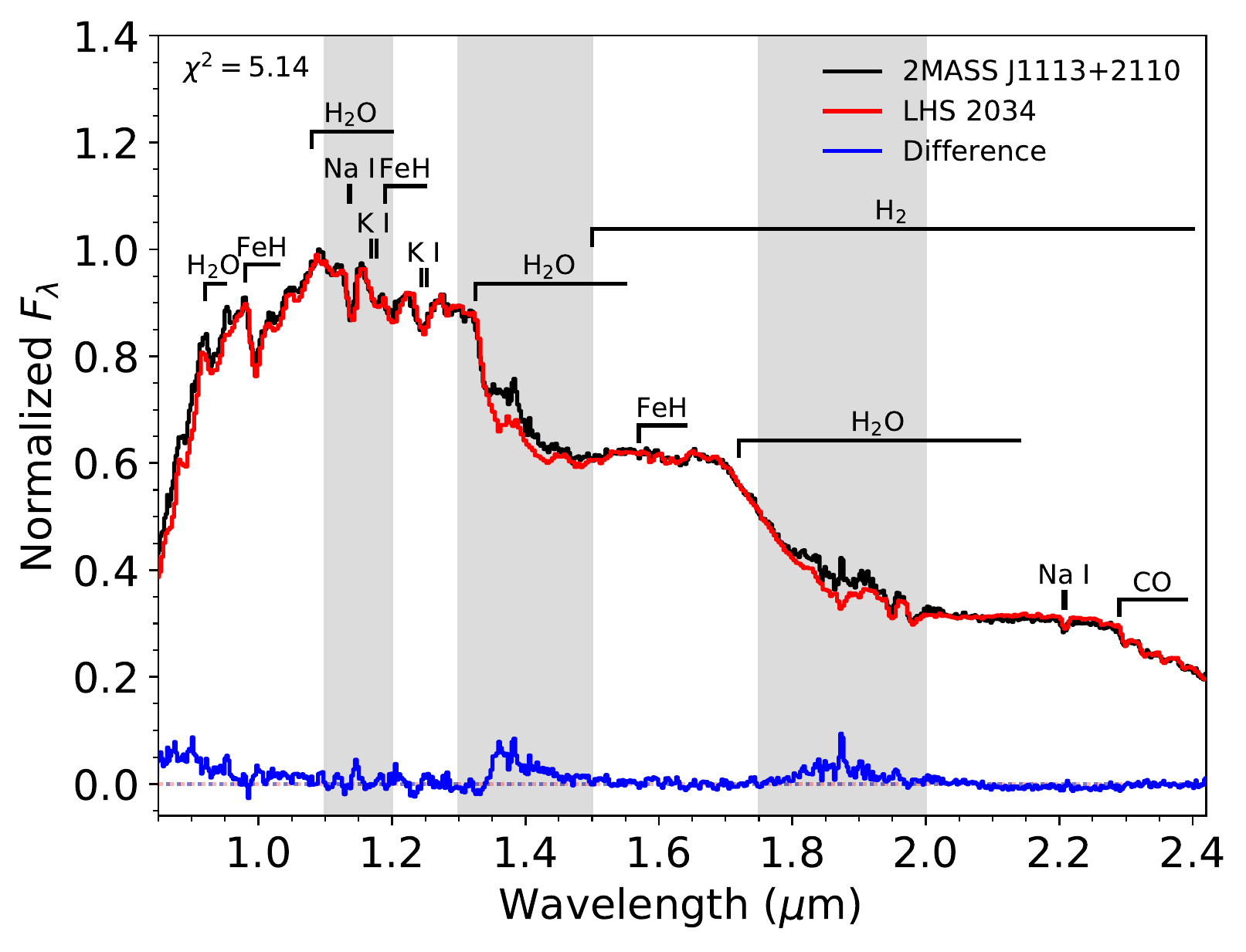}
\includegraphics[width=0.496\textwidth]{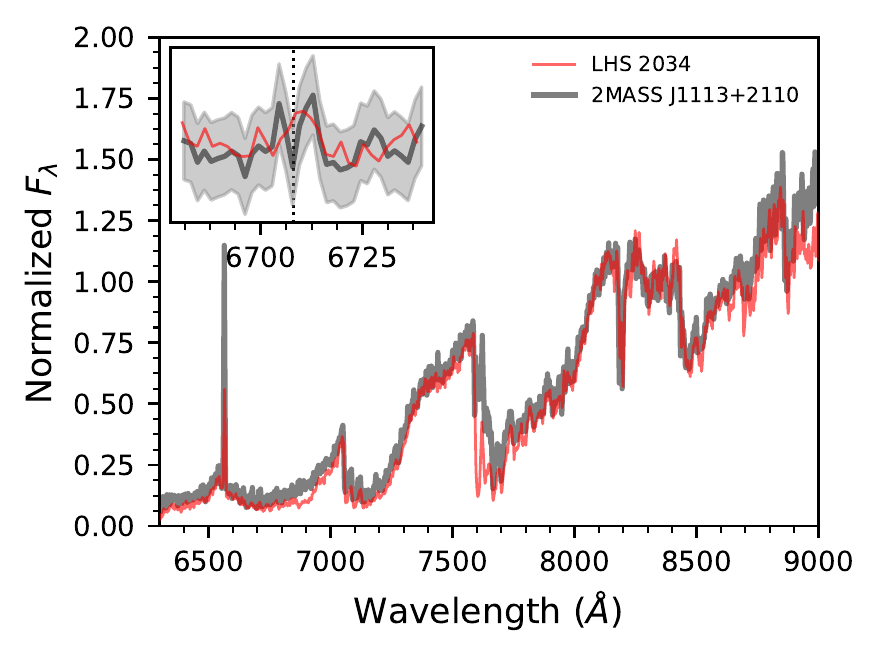}
\caption{Comparison spectra to 2MASS~J1113+2110 (black line), similar to Figure~\ref{fig:nir}.
\textit{Top}: Comparisons to the field M6 (left) and M7 (right) standards from \citet{kirkpatrick:2010:100}, fitting to only the 0.9--1.4~\um\ region. 
\textit{Bottom Left}: NIR comparison to LHS 2034, using spectra from \citet{bardalez-gagliuffi:2014:143}.
\textit{Bottom Right}: Optical comparison to LHS 2034, using spectra from \citet{reid:2007:2825}. The inset plot shows the location of the Li \textsc{i} feature (6708~\AA; black dotted line) and flux uncertainty (gray shaded region). 
\label{fig:companion}}
\end{figure*}

	The optical and NIR spectra are consistent with an M6 dwarf, albeit with a slightly redder NIR SED. As shown in Figure~\ref{fig:companion}, these data are particularly well matched to data for the M6 dwarf LHS 2034 \citep{shkolnik:2009:649, bardalez-gagliuffi:2014:143, newton:2014:20}. Although there are no obvious signatures of low-surface gravity in the NIR spectrum of 2MASS~J1113+2110 or LHS 2034, LHS 2034 has been previously proposed as a young star with an estimated age of $\sim$100~Myr based on He \textsc{i} (6678~\AA) emission and variable, strong H$\alpha$ emission \citep[EW$_{\mathrm{H}\alpha}=22.68$~\AA, $\Delta\mathrm{EW}_{\mathrm{H}\alpha}\approx10$~\AA;][]{shkolnik:2009:649}. 
	
	The H$\alpha$ equivalent width (EW$_{\mathrm{H}\alpha}=30\pm2$~\AA) of 2MASS~J1113+2110, is larger than 99.8\% of measured H$\alpha$ EWs from field M6 dwarfs in the SDSS spectroscopic M dwarf sample \citep{west:2011:97}, indicating that 2MASS~J1113+2110 is likely to be younger than average field objects. Only M6 dwarfs with ages $\lesssim100$~Myr tend to have such large H$\alpha$ EWs \citep{shkolnik:2009:649}. We searched for Li \textsc{i} (6708~\AA) absorption in the moderate-resolution spectrum (Figure~\ref{fig:companion}) but the S/N was too low to assert a robust detection. 
	
	We report the field-based photometric distance for 2MASS~J1113+2110 \citep[$54\pm9$~pc;][]{theissen:2017:92} and radius ($0.140\pm0.016~R_\odot$). This is primarily due to the ``young" calibration from \citet{filippazzo:2015:158} giving an estimate of $T_\mathrm{eff}\approx2298$~K, which is inconsistent with our MCMC value of $2767^{+42}_{-53}$~K. The field calibration from \citet{filippazzo:2015:158} and \citet{faherty:2016:10} gives $T_\mathrm{eff} \approx 2771$~K, which is more consistent with our value. Measured properties are reported in Table~\ref{tbl:candidate}.
	
	\citet{cook:2016:2192} flagged 2MASS~J1113+2110 as a potentially unresolved binary composed of an M dwarf and a lower-mass companion. The SpeX spectrum does not show evidence for a spectral binary \citep{bardalez-gagliuffi:2014:143}, although such sources generally have a T dwarf secondary; we cannot rule out a late-M or L dwarf companion. Additionally, the high-res NIRSPEC spectrum rules out a very close ($\lesssim$0.1~AU), double-lined spectroscopic binary. We obtained adaptive optics (AO) imaging of this object using the Near-infraRed Camera 2 (NIRC2) and laser guide star (LGS) AO system \citep{van-dam:2006:310, wizinowich:2006:297} on the Keck II 10-m Telescope on 2017 May 4 (UT). These data rule out companions more widely separated than $\gtrsim$0$\farcs$1 (projected separation $\gtrsim3$~AU) and brighter than $H\approx15.2$ \citep{bardalez-gagliuffi:2015:163}. While there is a small chance of a companion existing between these limits, it is unlikely that 2MASS~J1113+2110 is a close binary system.

	The spatial proximity, similar space motion, and evidence of youth in these two sources suggest a physical association even if not gravitationally bound. However, the large space density of low-mass stars and wide separation of the pair necessitate an investigation of the probability that these two objects are simply chance alignments with similar positions and kinematics. We assess the probability for a chance alignment using a method similar to \citet{dhital:2010:2566}, who simulated low-mass stellar populations and kinematics along a line-of-sight through the Galaxy to confirm wide, co-moving pairs. 
	
	To simulate star counts and kinematics along the line-of-sight of the binary, we used the Low-mass Kinematics model \citep[\textit{LoKi};][]{theissen:2017:165}, which builds stellar populations using an empirical luminosity function and empirical kinematic dispersions. We ran 10$^6$ simulations using a 3$^\circ$ field along the line of sight, integrated between 0--2500~pc, counting every simulation in which at least two stars were found within the projected separation of the candidate pair, and the distances and kinematics (proper motions and RVs) of the simulated stars were within the 1-$\sigma$ ranges for 2MASS~J1115+1937 and 2MASS~J1113+2110. We find the probability of chance alignment to be 1.8\%, making this a possible, but low probability, chance alignment.
		
	To further assess the physical association of this visual pair, we estimated the disruption timescale for a binary of this combined mass and separation as it orbits through the Galaxy. We estimated the average time for a binary with semi-major axis $a$ to be disrupted using Equation (18) from \citet[][based on the work of \citealt{weinberg:1987:367} and \citealt{close:2007:1492}]{dhital:2010:2566},
\begin{equation}
\label{eqn:time}
t_\ast \approx 1.212\, \frac{M_\mathrm{tot}}{a},
\end{equation}
where $a$ is the semi-major axis in pc, $t_\ast$ is the average lifetime of the binary in Gyr, and $M_\mathrm{tot}$ is the total mass of the binary in solar units. This equation assumes an average Galactic mass density of $0.11\, M_\odot\, \mathrm{pc}^{-3}$, and an average perturber with $V_\mathrm{rel}=20$ km s$^{-1}$, $M=0.7\, M_\odot$ (the average mass of a perturber), and a Coulomb logarithm of unity, $\Lambda=1$. To approximate the true semi-major axis, we used the statistical correction that accounts for the eccentricity and inclination angle of the binary orbit \citep[Equation 7 from][]{fischer:1992:178} to convert projected separation ($s$) into true separation ($a$),
\begin{equation}
\label{eqn:sep}
a \approx 1.26\, s = 1.26\, \Delta \theta\, d,
\end{equation}
where $d$ is the distance to the binary, and $\Delta\theta$ is the angular separation in radians. Using the uncertainty weighted average distance of the candidate pair ($d = 47\pm5$ pc), an uncertainty in $\Delta\theta$ of one-tenth of an SDSS pixel\footnote{\url{http://classic.sdss.org/dr3/instruments/imager/}} ($\sim$0.04\arcsec), and Equation~(\ref{eqn:sep}), we estimated $a=1.7\pm0.2$~pc. Using the maximum and minimum value for $a$ and the mass ranges from Table~\ref{tbl:candidate} in Equation~(\ref{eqn:time}) gave us a survival time between 32--75~Myr for this candidate pair. This indicates that its survival time may be on the order of the ages of the system components. If the co-moving nature of this system is confirmed in the future, through more precise proper motion measurements and a precise radial velocity measurement for 2MASS~J1115+1937, this may be an example of a co-ejected wide pair that has not had time to be dynamically dissolved through encounters in the Galactic field population \citep[e.g.,][]{caballero:2010:a98, oelkers:2017:259, oh:2017:257, price-whelan:2017:}.

\section{Assessing Kinematic Membership}
\label{membership}

	There are a number of known NYMGs with estimated distances similar to those of 2MASS~J1115+1937 and 2MASS~J1113+2110 \citep[for a detailed review see][]{mamajek:2016:21}. Several tools exist to assess membership in these NYMGs and include, but are not limited to, the Bayesian Analysis for Nearby Young AssociatioNs (BANYAN) I \citep{malo:2013:88}, II \citep{gagne:2014:121}, and $\Sigma$ (Gagn\'e et al. 2017, \textit{ApJS}, submitted), the LocAting Constituent mEmbers In Nearby Groups (LACEwING) \citep{riedel:2017:95}, and the convergent point analysis tool of \citet{rodriguez:2013:101}. 
	
	The convergent point analysis tool uses positions and proper motions to trace back the tangential motions and match to the convergent point of a kinematic group. Using the analysis tool of \citet{rodriguez:2013:101}, we find that there is a high probability that 2MASS~J1115+1937 and 2MASS~J1113+2110 could be members of multiple NYMGs, but with drastically different target radial velocities (RVs) and distances. The BANYAN I, II, and $\Sigma$ tools and LACEwING use positions, proper motions, RVs, and distances to accurately calculate 3-D kinematics and better assess group membership. The RV for 2MASS~J1113+2110 was measured using the NIRSPEC data and the forward-modeling method described in \citet{burgasser:2015:104} and \citet{blake:2010:684}. All four methods give $<0.001\%$ probability that either 2MASS~J1115+1937 or 2MASS~J1113+2110 are members of any of the NYMGs tested in these tools\footnote{
	(1) $\epsilon$ Cham$\ae$leontis; 
	(2) $\eta$ Cham$\ae$leontis; 
	(3) TW Hydrae; 
	(4) $\beta$ Pictoris; 
	(5) 32 Orionis; 
	(6) Octans; 
	(7) Tucana-Horologium; 
	(8) Columba; 
	(9) Carina; 
	(10) Argus; 
	(11) AB Doradus;
	(12) Carina;
	(13) Carina-Near; 
	(14) Coma Berenices; 
	(15) Ursa Major; 
	(16) $\rchi^{01}$ Fornax; 
	(17) Hyades;
	(18) 118 Tau;
	(19) Corona Australis;
	(20) Lower Centaurus-Crux;
	(21) Platais 8;
	(22) Pleiades;
	(23) $\rho$ Ophiuci;
	(24) IC 2602;
	(25) IC 2391;
	(26) Upper Centaurus-Lupus;
	(27) Upper Corona Australis; and
	(28) Upper Scorpius.}.

\section{Discussion}
\label{discussion}
	
	2MASS~J1115+1937 joins a growing group of young, isolated low-mass stars and brown dwarfs \citep[e.g.,][]{cruz:2009:3345, gagne:2015:33, faherty:2016:10}. Although \citet{allers:2013:79} show that low gravity classification and redness (using NIR colors) of an object do not necessarily prove youth, the elevated near-UV continuum, hydrogen and helium emission, and MIR excess of 2MASS~J1115+1937 are all consistent with an age younger than 45~Myr. 2MASS~J1115+1937 may be a member of a kinematic association awaiting discovery, or the result of an ejection from a young association. This potentially makes 2MASS~J1115+1937 a very important benchmark for brown dwarf formation scenarios, and possibly a new optical, spectroscopic standard.
	
	2MASS~J1113+2110 may be associated with 2MASS J1115+1937. The strong H$\alpha$ emission is consistent with a relatively young field star ($\lesssim$100~Myr). While the NIR spectrum of 2MASS~J1113+2110 is not classified as low gravity, it falls within a spectral type regime where gravity classifications converge \citep[e.g., see Figures 20, 22, and 24 from][]{allers:2013:79}. A trigonometric parallax measurement would help associate or dissociate 2MASS~J1113+2110 with 2MASS~J1115+1937. Unfortunately, both sources are too faint to be detected by \textit{Gaia}, and must wait for a future astrometric measurement. Detection of Li \textsc{i} with a high-resolution optical spectrum would also constrain the mass to $< 0.6 M_\odot$, and the age to $< 100$ Myr, providing an independent check on the age of 2MASS~J1115+1937.

\acknowledgments

	The authors would like to thank the helpful suggestions by the anonymous referee which contributed greatly to the quality of this manuscript. A.J.B. acknowledges funding support from the NSF under award No. AST-1517177 and the US-UK Fulbright Commission. This material is based upon work supported by NASA under Grant No. NNX15AI75G and Grant No. NNX16AF47G issued through the Astrophysics Data Analysis Program. 

The authors recognize and acknowledge the very significant
cultural role and reverence that the summit of
Mauna Kea has always had within the indigenous Hawaiian
community. We are most fortunate and grateful to
have the opportunity to conduct observations from this
mountain.
		
Funding for SDSS-IV has been provided by
the Alfred P. Sloan Foundation, the U.S. Department of Energy Office of
Science, and the Participating Institutions. SDSS-IV acknowledges
support and resources from the Center for High-Performance Computing at
the University of Utah. The SDSS web site is \url{www.sdss.org}.

SDSS-IV is managed by the Astrophysical Research Consortium for the 
Participating Institutions of the SDSS Collaboration including the 
Brazilian Participation Group, the Carnegie Institution for Science, 
Carnegie Mellon University, the Chilean Participation Group, the French Participation Group, Harvard-Smithsonian Center for Astrophysics, 
Instituto de Astrof\'isica de Canarias, The Johns Hopkins University, 
Kavli Institute for the Physics and Mathematics of the Universe (IPMU) / 
University of Tokyo, Lawrence Berkeley National Laboratory, 
Leibniz Institut f\"ur Astrophysik Potsdam (AIP), 
Max-Planck-Institut f\"ur Astronomie (MPIA Heidelberg), 
Max-Planck-Institut f\"ur Astrophysik (MPA Garching), 
Max-Planck-Institut f\"ur Extraterrestrische Physik (MPE), 
National Astronomical Observatory of China, New Mexico State University, 
New York University, University of Notre Dame, 
Observat\'ario Nacional / MCTI, The Ohio State University, 
Pennsylvania State University, Shanghai Astronomical Observatory, 
United Kingdom Participation Group,
Universidad Nacional Aut\'onoma de M\'exico, University of Arizona, 
University of Colorado Boulder, University of Oxford, University of Portsmouth, 
University of Utah, University of Virginia, University of Washington, University of Wisconsin, 
Vanderbilt University, and Yale University.
 
	This publication makes use of data products from 2MASS, which is a joint project of the University of Massachusetts and the IPAC/Caltech, funded by NASA and NSF. This publication also makes use of data products from the \textit{Wide-field Infrared Survey Explorer}, which is a joint project of UCLA, and JPL/Caltech, funded by NASA. 
	
	This research has benefitted from the SpeX Prism Libraries, maintained by Adam Burgasser at \url{http://pono.ucsd.edu/~adam/browndwarfs/spexprism}; and the M, L, and T dwarf compendium housed at \url{http://DwarfArchives.org} and maintained by Chris Gelino, Davy Kirkpatrick, and Adam Burgasser. This research has also made use of the SIMBAD database and the VizieR catalogue access tool, operated at CDS, Strasbourg, France; the NASA/ IPAC Infrared Science Archive, which is operated by the Jet Propulsion Laboratory, California Institute of Technology, under contract with the National Aeronautics and Space Administration; NASA's Astrophysics Data System; and Astropy, a community-developed core Python package for Astronomy \citep{astropy-collaboration:2013:a33}. Plots in this publication were made using Matplotlib \citep{hunter:2007:90}. This research has made use of the SIMBAD database, operated at CDS, Strasbourg, France \citep{wenger:2000:9}. 
	
\facilities{IRTF (SpeX), Keck:II (NIRC2, NIRSPEC), DCT (DeVeny spectrograph), IRSA, \textit{WISE}}.

\software{SpeXtool \citep{vacca:2003:389, cushing:2004:362}, SPLAT (Burgasser et al., in preparation), Astropy \citep{astropy-collaboration:2013:a33}, Matplotlib \citep{hunter:2007:90}, BANYAN $\Sigma$ (Gagn\'e et al. 2017, \textit{ApJS}, submitted), LACEwING \citep{riedel:2017:95}, BANYAN II \citep{gagne:2014:121}, BANYAN I \citep{malo:2013:88}, the convergent point tool \citep{rodriguez:2013:101}, emcee \citep{foreman-mackey:2013:306}, LoKi \citep{theissen:2016:a}, Sublime Text}.

\bibliography{arxiv.bbl}
\bibliographystyle{aasjournal}

\end{document}